# Quarkonium Masses in the N-dimensional Space Using the Analytical Exact Iteration Method

E. M. Khokha [1], M. Abu-Shady [2], and T. A. Abdel-Karim [2]

[1]Department of Basic Science, Modern Academy for Engineering and Technology, Cairo, Egypt

[2]Department of Applied Mathematics, Faculty of Science, Menoufia University, Shebin El- Kom Egypt

**Email address:**

emad_mohamed_2015@yahoo.com(E. M. Khokha), dr.abushady@gmail.com (M. Abu-Shady), tabdelkarim63@yahoo.com (T. A. Abdel-Karim)

**Abstract:** The $N$- dimensional radial Schrödinger equation with an extended Cornell potential is solved. The analytical exact iteration method (AEIM) is applied. The energy eigenvalues are calculated in the $N$–dimensional space. The charmonium meson, the bottomonium meson and the $b\bar{c}$ meson masses are calculated in the $N$-dimensional space. The special cases are obtained from the general case. The study of the effect of dimensionality number is studied. The mean value of the radius and the mean square velocity of charmonium meson, bottomonium meson, and $b\bar{c}$ meson are calculated. The present results are improved in comparison with other recent studies and are in good agreement with the experimental data. Therefore, the present method with the present potential gives successfully description of heavy quarkonium properties.

## 1. Introduction

Studying heavy quarkonium systems such as charmonium meson, bottomonium meson, and $b\bar{c}$ meson have a vital role for comprehension the quantitative tests of (QCD) and the standard model [1]. The study of these systems is very important. A lot of heavy quarkonium systems can be studied within the Schrödinger equation [2].

The solution of Schrödinger equation with spherically symmetric potential is one of the important problems in physics. It plays a large vital role for spectroscopy, atoms, molecules, and nuclei, in particular, the properties of constituents particles and dynamics of their interactions.

Several papers have focused on the study of multi-dimensional space or $N$-dimensional space [3]. The study of multi-dimensional space problems is more general, and it can easily obtain the results in the lower dimensional space [4]. The effect of the $N$-dimensional space is studied on the energy levels of quantum mechanical systems [5]. The Hydrogen atom [6] and the harmonic oscillator [7] have been studied in the $N$-dimensional space. In addition, other potentials have been solved in the N-dimensional space such as Coulomb potential [8], Psedoharmonic potential [9], Kratzer potential [10], Hulthen potential [11], Poschl-Teller potential [12], Mie-type potential [13], energy dependent potential [14], forth-order inverse power potential [15], Hua potential [16], global potential [17], and Cornell potential [18].

Schrödinger equation has been solved by using numerous methods such as quasi-linearization method (QLM) [20], Hill determinant method [21], point canonical transformation (PCT) [22], and numerical methods [23-25]. The $N$-dimensional Schrödinger equation is solved by the Nikiforov-Uvarov (UV) method [5, 11, 14], (AIM) method [2], Laplace Transform method [3, 8], (SUSQM) method [26], power series technique [27], and Pekeris type approximation [16, 28].

In this paper, we consider the confining potential is the extended Cornell potential [19].

$$V(r) = ar^2 + br - \frac{c}{r} - \frac{d}{r^2} \qquad (1)$$

The aim of this paper is to calculate the solutions of the $N$-radial Schrodinger equation using the analytical exact iteration method (AEIM) [19, 29, 30]. Additionally, the present results are applied on quarkonium properties in comparison with experimental data and other recent studies.

The paper is organized as follows: In Sec. 2, the exact solution of the $N$-dimensional radial Schrödinger equation is derived. In Sec.3, the results are discussed. In Sec. 4, summary and conclusion are presented.

## 2. Exact Solution of the *N*-dimensional Radial Schrödinger Equation With the Extended Cornell Potential

The $N$-dimensional radial Schrödinger equation for two particles interacting via symmetric potential (1) takes the form [2].

$$\left[\frac{d^2}{dr^2} + \frac{N-1}{r}\frac{d}{dr} - \frac{l(l+N-2)}{r^2} + 2\mu\left(E_{nl} - V(r)\right)\right]\psi_n(r) = 0 \qquad (2)$$

where l, N and μ denote the angular quantum number, the dimensional number, and μ is the reduced mass of the two particles $\mu = \frac{m_q m_{\bar{q}}}{m_q + m_{\bar{q}}}$, respectively.

Inserting the transformation

$$\psi_n(r) = \frac{1}{r^{\frac{(N-1)}{2}}} \varphi_n(r). \tag{3}$$

Eq. (2) takes the form

$$\varphi_n''(r) + \left[2\mu\left(E_{nl} - ar^2 - br + \frac{c}{r} + \frac{d}{r^2}\right) - \frac{\frac{1}{4}(N^2 - 4N + 3) + l(l+N-2)}{r^2}\right]\varphi_n(r) = 0. \tag{4}$$

Using the simplifications:

$$\varepsilon_{nl} = 2\mu E_{nl}, a_1 = 2\mu a, b_1 = 2\mu b, c_1 = 2\mu c, d_1 = 2\mu d. \tag{5}$$

Eq. (4) reduces to

$$\varphi_n''(r) + \left[\varepsilon_{nl} - a_1 r^2 - b_1 r + \frac{c_1}{r} + \frac{d_1 - \frac{1}{4}(N^2 - 4N + 3) - l(l+N-2)}{r^2}\right]\varphi_n(r) = 0. \tag{6}$$

To solve the *N*-dimensional radial Schrödinger equation with the analytical exact iteration method (AEIM). The following ansatz for the wave function is assumed [19].

$$\varphi_n(r) = f_n(r) \exp[g(r)]. \tag{7}$$

Where

$$f_n(r) = \begin{cases} 1, & n = 0, \\ \prod_{i=1}^{n}(r - \alpha_i^{(n)}), & n = 1,2,\dots, \end{cases} \tag{8a}$$

$$g(r) = -\frac{1}{2}\alpha r^2 - \beta r + \delta \ln r, \alpha > 0, \beta > 0. \tag{8b}$$

From Eq. (7), we obtain

$$\varphi_n''(r) = \left[g''(r) + g'^2(r) + \frac{f_n''(r) + 2f_n'(r) g(r)}{f(r)}\right]\varphi_n(r). \tag{9}$$

Comparing Eqs. (6) and (9) yields

$$a_1 r^2 + b_1 r - \frac{c_1}{r} + \frac{\frac{1}{4}(N^2 - 4N + 3) + l(l+N-2) - d_1}{r^2} - \varepsilon_{nl} = g''(r) + g'^2(r) + \frac{f_n''(r) + 2f_n'(r) g(r)}{f(r)}. \tag{10}$$

At *n*=0, By substituting Eqs. (8) into Eq. (10) gives

$$a_1 r^2 + b_1 r - \frac{c_1}{r} + \frac{\frac{1}{4}(N^2 - 4N + 3) + l(l+N-2) - d_1}{r^2} - \varepsilon_{0l} = \alpha^2 r^2 + 2\alpha\beta r - \alpha[1 + 2(\delta + 0)] + \beta^2 - \frac{2\beta\delta}{r} + \frac{\delta(\delta - 1)}{r^2}. \tag{11}$$

By comparing the corresponding powers of *r* on both sides of Eq. (11), one obtains

$$\alpha = \sqrt{a_1}, \tag{12a}$$

$$\beta = \frac{b_1}{2\sqrt{a_1}}, \tag{12b}$$

$$c_1 = 2\beta(\delta + 0), \tag{12c}$$

$$\delta = \frac{1}{2}(1 \pm \hat{l}), \hat{l} = \sqrt{(N-2)^2 + 4l(l+N-2) - 8\mu d}, \tag{12d}$$

$$\varepsilon_{0l} = \alpha[1 + 2(\delta + 0)] - \beta^2. \tag{12e}$$

To get well-behaved solutions at the origin and the infinity, $\delta$ is taken in the following from Eq. (12d) as

$$\delta = \frac{1}{2}\left(1 + \sqrt{(N-2)^2 + 4l(l+N-2) - 8\mu d}\right). \tag{13}$$

From Eqs. (5), (12a), (12e) and (13). The ground state energy is:

$$E_{0l} = \sqrt{\frac{a}{2\mu}}(2 + \hat{l}) - \frac{2\mu c^2}{(1+\hat{l})^2}. \tag{14}$$

Where the coefficient c satisfy this restriction

$$c = \frac{b}{2}(1 + \hat{l})\sqrt{\frac{1}{2\mu a}}. \tag{15}$$

Substitution of, $\alpha$, $\beta$ and $\delta$ from Eqs. (12a), (12b) and (13) together with Eqs. (3), (7) and (8) gives the following ground state wave function:

$$\psi_0(r) = N_0 r^{\frac{(\hat{l}-N+2)}{2}} \exp\left[-\frac{1}{2}\sqrt{2\mu a}\, r^2 - \frac{2\mu c}{(1+\hat{l})} r\right]. \tag{16}$$

Secondly, for the first node (*n*=1), we use f(r) = r-$\alpha_1^{(1)}$ and g(r) from Eq. (8b) to solve Eq. (10) gives:

$$a_1 r^2 + b_1 r - \frac{c_1}{r} + \frac{\frac{1}{4}(N^2 - 4N + 3) + l(l+N-2) - d_1}{r^2} - \varepsilon_{1l} = \alpha^2 r^2 + 2\alpha\beta r - \alpha[1 + 2(\delta + 1)] + \beta^2 - \frac{2[\beta(\delta+1) + \alpha\alpha_1^{(1)}]}{r} + \frac{\delta(\delta-1)}{r^2}. \tag{17}$$

Then, the relations between the potential parameters and the coefficients $\alpha, \beta, \delta$ and $\alpha_1^{(1)}$ are given by:

$$\alpha = \sqrt{a_1}, \tag{18a}$$

$$\beta = \frac{b_1}{2\sqrt{a_1}}, \tag{18b}$$

$$c_1 = 2\beta(\delta + 0), \tag{18c}$$

$$\varepsilon_{1l} = \alpha[1 + 2(\delta + 1)] - \beta^2, \tag{18d}$$

$$c_1 - 2\beta(\delta + 1) = 2\alpha\alpha_1^{(1)}, \tag{18e}$$

$$(c_1 - 2\beta\delta)\alpha_1^{(1)} = 2\delta, \tag{18f}$$

$$d_1 = \frac{1}{4}(N^2 - 4N + 3) + l(l+N-2) - \delta(\delta-1), \tag{18g}$$

Where $c, \alpha_1^{(1)}$ are obtained from the constraint relations,

$$c = \frac{b}{2\sqrt{2\mu a}}(3+\hat{l}) + 2\sqrt{\frac{a}{2\mu}}\alpha_1^{(1)}, \tag{19a}$$

$$\alpha\alpha_1^{(1)^2} + \beta\alpha_1^{(1)} - \delta = 0, \tag{19b}$$

So, one obtain

$$\alpha_1^{(1)} = -\frac{b}{4a} \pm \sqrt{\frac{b^2}{16a^2} + \frac{(1+\hat{l})}{\sqrt{8\mu a}}}, \tag{19c}$$

Then, the energy eigenvalue is:

$$E_{1l} = \sqrt{\frac{a}{2\mu}}(4+\hat{l}) - \frac{b^2}{4a}. \tag{20}$$

The corresponding wave function is:

$$\psi_1(r) = N_1(r - \alpha_1^{(1)}) r^{\frac{(\hat{l}-N+2)}{2}} \exp\left[-\frac{1}{2}\sqrt{2\mu a}\, r^2 - \sqrt{\frac{\mu}{2a}}\, b\, r\right]. \tag{21}$$

The second node (n=2), we use $f(r) = (r-\alpha_1^{(2)})(r-\alpha_2^{(2)})$ and $g(r)$ from Eq. (8b) to solve Eq. (10) gives:

$$a_1 r^2 + b_1 r - \frac{c_1}{r} + \frac{\frac{1}{4}(N^2-4N+3)+l(l+N-2)-d_1}{r^2} - \varepsilon_{2l} = \alpha^2 r^2 + 2\alpha\beta r - \alpha[1+2(\delta+2)] + \beta^2 - \frac{2[\beta(\delta+2)+\alpha\sum_{i=1}^{2}\alpha_i^{(2)}]}{r} + \frac{\delta(\delta-1)}{r^2}. \tag{22}$$

Therefore, the relations between the potential parameters and the coefficients, $\beta, \delta, \alpha_1^{(2)}$ and $\alpha_2^{(2)}$ are given by:

$$\alpha = \sqrt{a_1}, \tag{23a}$$

$$\beta = \frac{b_1}{2\sqrt{a_1}}, \tag{23b}$$

$$\delta = \frac{1}{2}(1+\hat{l}), \tag{23c}$$

$$\varepsilon_{2l} = \alpha[1+2(\delta+2)] - \beta^2, \tag{23d}$$

$$c_1 - 2\beta(\delta+2) = 2\alpha\sum_{i=1}^{2}\alpha_i^{(2)}, \tag{23e}$$

$$(c_1 - 2\beta\delta)\sum_{i<j}^{2}\alpha_i^{(2)}\alpha_j^{(2)} = 2\delta\sum_{i=1}^{2}\alpha_i^{(2)}, \tag{23f}$$

$$[c_1 - 2\beta(\delta+1)]\sum_{i=1}^{2}\alpha_i^{(2)} = 4\alpha\sum_{i<j}^{2}\alpha_i^{(2)}\alpha_j^{(2)} + 2(2\delta+1). \tag{23g}$$

The coefficients $\alpha_1^{(2)}$ and $\alpha_2^{(2)}$ are given by the constraint relations,

$$\alpha\sum_{i=1}^{2}\alpha_i^{(2)^2} + \beta\sum_{i=1}^{2}\alpha_i^{(2)} - (2\delta+1) = 0. \tag{24a}$$

$$\delta\sum_{i=1}^{2}\alpha_i^{(2)^2} - \left(\beta\sum_{i=1}^{2}\alpha_i^{(2)} + 1\right)\sum_{j<k}^{2}\alpha_j^{(2)}\alpha_k^{(2)} - 2\alpha\sum_{j<k}^{2}\alpha_i^{(2)^2}\alpha_k^{(2)^2} = 0. \tag{24b}$$

Then, the energy eigenvalue is:

$$E_{2l} = \sqrt{\frac{a}{2\mu}}(6+\hat{l}) - \frac{b^2}{4a}. \tag{25}$$

The corresponding wave function is:

$$\psi_2(r) = N_2 \prod_{i=1}^{2}(r - \alpha_i^{(2)})\, r^{\frac{(\hat{l}-N+2)}{2}} \exp\left[-\frac{1}{2}\sqrt{2\mu a}\, r^2 - \sqrt{\frac{\mu}{2a}}\, b\, r\right]. \tag{26}$$

Similarly, the iteration method is repeated many times. Thus, the exact energy formula for the confining potential (1) with any arbitrary $n$ state is obtained:

$$E_{nl} = \sqrt{\frac{a}{2\mu}}(2+2n+\hat{l}) - \frac{b^2}{4a}. \tag{27}$$

The corresponding wave functions for any $n$ state are:

$$\psi_n(r) = N_v \prod_{i=1}^{n}(r - \alpha_i^{(n)})\, r^{\frac{(\hat{l}-N+2)}{2}} \exp\left[-\frac{1}{2}\sqrt{2\mu a}\, r^2 - \sqrt{\frac{\mu}{2a}}\, b\, r\right]. \tag{28}$$

where the relations between the parameters of potential and the coefficients $\alpha, \beta, \delta, \alpha_1^{(n)}, \alpha_1^{(n)}, \alpha_2^{(n)}, \ldots, \alpha_3^{(n)}$ are

$$\alpha = \sqrt{a_1},\ \beta = \frac{b_1}{2\sqrt{a_1}},\ \delta = \frac{1}{2}(1+\hat{l}),$$

$$\varepsilon_{nl} = \alpha[1+2(\delta+n)] - \beta^2, \tag{29a}$$

$$c_1 - 2\beta(\delta+n) = 0, (n=0), \tag{29b}$$

$$c_1 - 2\beta(\delta+n) = 2\alpha\sum_{i=1}^{n}\alpha_i^{(n)}, n=1,2,3,\ldots,$$

$$[c_1 - 2\beta(\delta+n-1)]\sum_{i=1}^{n}\alpha_i^{(n)} = n(2\delta+n-1), (n=1) \tag{29c}$$

$$[c_1 - 2\beta(\delta+n-1)]\sum_{i=1}^{n}\alpha_i^{(n)} = 4\alpha\sum_{i<j}^{n}\alpha_i^{(n)}\alpha_j^{(n)} + n(2\delta+n-1), n=2,3,4,\ldots,$$

$$[c_1 - 2\beta(\delta+n-2)]\sum_{i<j}^{n}\alpha_i^{(n)}\alpha_j^{(n)} = (n-1)(2\delta+n-2)\sum_{i=1}^{n}\alpha_i^{(n)}, (n=2), \tag{29d}$$

$$[c_1 - 2\beta(\delta+n-2)]\sum_{i<j}^{n}\alpha_i^{(n)}\alpha_j^{(n)} = (n-1)(2\delta+n-2)\sum_{i<j}^{n}\alpha_i^{(n)}\alpha_j^{(n)} + 4\alpha\sum_{i<j<k}^{n}\alpha_i^{(n)}\alpha_j^{(n)}\alpha_k^{(n)}, n=3,4,5,\ldots,$$

$$(c_1 - 2\beta\delta)\sum_{i<j<k}^{n}\alpha_i^{(n)}\alpha_j^{(n)}\alpha_k^{(n)} = 2\delta\sum_{i<j}^{n}\alpha_i^{(n)}\alpha_j^{(n)}, (n=3), \tag{29e}$$

And so on

Finally, from Eqs. (27) and (12d) together we can write the energy eigenvalue of Eq. (4) in $N$-dimensional as

$$E_{nl}^N = \sqrt{\frac{a}{2\mu}}\left(2n+2\pm\sqrt{(N-2)^2+4l(l+N-2)-8\mu d}\right) - \frac{b^2}{4a}. \tag{30}$$

# 3. Discussion of Results

## 3.1. Mass Spectra of Heavy Quarkonia in the N-dimensional Space

In this section, the properties of charmonuim meson, bottomonium meson, and $b\bar{c}$ meson are calculated, in which the quarkonium meson have quark and antiquark masses. The following relation [1] is used for determining quarkonium masses in the *N*-dimensional space

$$M = m_q + m_{\bar{q}} + E_{nl}^N \qquad (31)$$

By substituting Eq. (30) into Eq. (31), the quarkonium mass in the N-dimensional space takes the following form

$$M = m_q + m_{\bar{q}} + \sqrt{\frac{a}{2\mu}}\left(2n + 2 \pm \sqrt{(N-2)^2 + 4l(l+N-2) - 8\mu d}\right) - \frac{b^2}{4a}. \qquad (32)$$

Thus, the charmonium mass is given by:

$$M_c = 2m_c + \sqrt{\frac{a}{m_c}}\left(2n + 2 \pm \sqrt{(N-2)^2 + 4l(l+N-2) - 4m_c d}\right) - \frac{b^2}{4a}. \qquad (33)$$

Similarly, the bottomonium mass is calculated from Eq. (33) by replacing $m_c$ by $m_b$.

**Table 1.** Mass spectra of charmonuim for parameters ($m_c$= 1.209 GeV, a = 0.0673 GeV$^3$, b = 0.0895 GeV$^2$, d = 0.001 GeV$^{-1}$)

| State | P. Work | Exp. | [27] | [2] | [4] | N=4 | N=5 | N=6 |
|---|---|---|---|---|---|---|---|---|
| 1S | 3.095481 | 3.096 | 3.078 | 3.078 | 3.096 | 3.331703 | 3.567735 | 3.803718 |
| 1P | 3.567735 | - | 3.415 | 3.415 | 3.433 | 3.803718 | 4.039683 | 4.275638 |
| 1D | 4.039683 | - | 3.581 | 4.187 | 3.686 | 4.275638 | 4.511588 | 4.747534 |
| 2S | 3.567354 | 3.686 | 3.749 | 3.752 | 3.767 | 3.803575 | 4.039607 | 4.27559 |
| 2P | 4.039607 | 3.773 | 3.917 | 4.413 | 3.910 | 4.27559 | 4.511555 | 4.74751 |
| 3S | 4.039226 | 4.040 | 4.085 | 5.297 | 3.984 | 4.275448 | 4.511479 | 4.747463 |
| 4S | 4.511098 | 4.263 | 4.589 | 6.407 | 4.150 | 4.74732 | 4.983351 | 5.219335 |

**Table 2.** Mass spectra of bottomonium for parameters ($m_b$=4.823 GeV, a= 0.09347 GeV$^3$, b= 0.34457 GeV$^2$, d= 0.001 GeV$^{-1}$)

| State | P. Work | Exp. | [27] | [2] | [4] | N=4 | N=5 | N=6 |
|---|---|---|---|---|---|---|---|---|
| 1S | 9.74473 | 9.46 | 9.510 | 9.510 | 9.460 | 9.884619 | 10.02406 | 10.16338 |
| 1P | 10.02406 | - | 9.862 | 9.862 | 9.840 | 10.16338 | 10.30266 | 10.44192 |
| 1D | 10.30266 | - | 10.038 | 10.627 | 10.023 | 10.44192 | 10.58116 | 10.7204 |
| 2S | 10.02315 | 10.023 | 10.214 | 10.214 | 10.140 | 10.16304 | 10.30248 | 10.4418 |
| 2P | 10.30248 | - | 10.390 | 10.944 | 10.160 | 10.4418 | 10.58108 | 10.72034 |
| 3S | 10.30158 | 10.355 | 10.566 | 11.726 | 10.280 | 10.44147 | 10.5809 | 10.72023 |
| 4S | 10.580 | 10.580 | 11.094 | 12.834 | 10.420 | 10.71989 | 10.85933 | 10.99865 |

**Table 3.** Mass spectra of $b\bar{c}$ meson for parameters ($m_c$=1.209 GeV, $m_b$=4.823 GeV, $a$= 0.2793 GeV$^3$, b= 0.999 GeV$^2$, d= 0.001 GeV$^{-1}$)

| State | P. Work | Exp. | [32] | [33] | [34] | N=4 | N=5 | N=6 |
|---|---|---|---|---|---|---|---|---|
| 1S | 6.277473 | 6.277 | 6.349 | 6.264 | 6.270 | 6.658294 | 7.038623 | 7.418830 |
| 1P | 7.038623 | - | 6.715 | 6.700 | 6.699 | 7.418830 | 7.798987 | 8.179120 |
| 2S | 7.037641 | - | 6.821 | 6.856 | 6.835 | 7.418462 | 7.798791 | 8.178997 |
| 2P | 7.798791 | - | 7.102 | 7.108 | 7.091 | 8.178997 | 8.559155 | 8.939287 |
| 3S | 7.797808 | - | 7.175 | 7.244 | 7.193 | 8.178630 | 8.558959 | 8.939165 |

We choose the positive sign in Eq. (33) in the present computations as in Refs. [4, 31]. The free parameters of the present computations a, b and d are fitted with experimental data and Eq. (33). Ikhdair and Hamzavi [19] solved Schrodinger equation in three dimensions for the same potential of the present work and calculated the energy eigenvalues. It can easily obtain their results if we put (*N*=3) in Eq. (30). In addition, we apply our results to quarkonium properties. In Refs. [2, 27], the potential is a special case of the present potential when we put (*d = 0*). Also, In Ref. [4], the potential is a special case of the present potential when setting (*b=0* and *d = 0*).

In Table 1, the charmonium mass is computed from Eq. (33) in comparison with other theoretical studies [2, 4, 27], and the experimental data. In Ref. [27], the authors computed the charmonium and bottomonium masses using the power series technique for the quark-antiquark interaction potential. In Ref. [2], the authors computed the charmonium and bottomonium masses using the asymptotic iteration method (AIM) for the quark-antiquark interaction potential. In Ref. [4], the authors computed the charmonium and bottomonium masses using the Nikiforov-Uvarov (UV) method. The present results are in good agreement with experimental data and are improved in comparison with results in Refs. [2, 4, 27]. In Table 2, the bottomonium mass is computed. Most states of bottomonium are improved in comparison with other theoretical studies [2, 4, 27], and are in good agreement with experimental data. In Table 3, the $b\bar{c}$ meson mass is computed. The 1S state is closed with experimental data and the other states are in good agreement in comparison with Refs. [32, 33, 34]. The charmonium mass, the bottomonium mass, and $b\bar{c}$ mass increase with increasing dimensional number due to an increase in the binding energy.

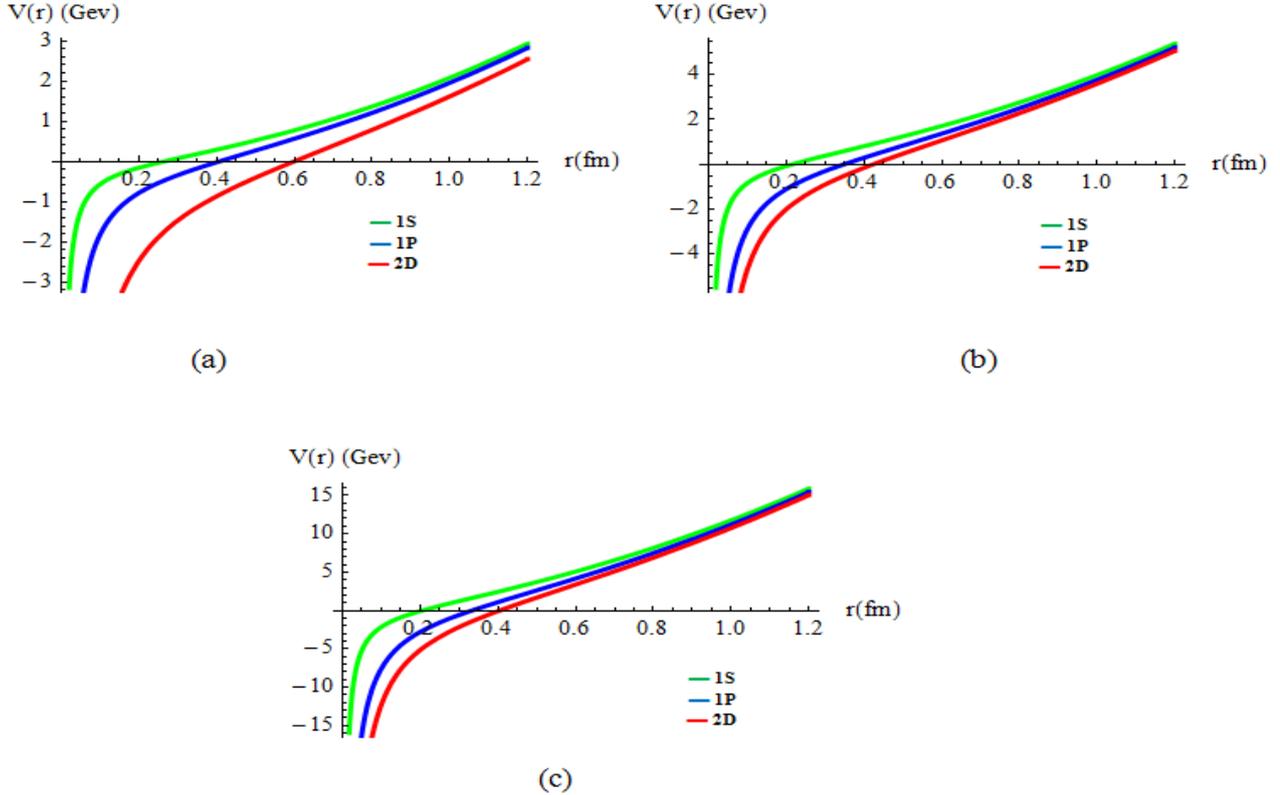

***Fig 1.*** *The present potential is plotted as a function of distance (r) for different states, where (a) represents charmonium states, (b) represents bottomonium states, and (c) represents $b\bar{c}$ meson states.*

In Fig. 1**.** The present potential is plotted as a function of distance (*r*) for different states for the range of distances 0.1 fm $\leq r \leq$ 1 fm as in Ref. [35]. The three curves increase with increasing distance (*r*). Additionally, all curves shift to upper values by increasing angular quantum number. This behavior is in agreement with Ref. [35].

## 4. Radius of Bound State

The mean radius $\langle r \rangle$ and the mean square velocity $\langle v^2 \rangle$ of the bound states of charmonium meson, bottomonium mseon and $b\bar{c}$ meson are calculated by using the Virial theorem [35]. The relation of the mean-kinetic energy and the potential energy is given by:

$$\langle T \rangle = \frac{1}{2} \langle r \frac{dV}{dr} \rangle. \tag{34}$$

The present potential is $V = ar^2 + br - \frac{c}{r} - \frac{d}{r^2}$, reduces

$$E = 2a\langle r^2 \rangle + \frac{3b\langle r \rangle}{2} - \frac{c}{2\langle r \rangle}, \tag{35}$$

The mass relation

$$M = 2m + 2a\langle r^2 \rangle + \frac{3b\langle r \rangle}{2} - \frac{c}{2\langle r \rangle}. \tag{36}$$

The mean square velocity

$$\langle v^2 \rangle = \frac{2}{M}\left(a\langle r^2 \rangle + \frac{b\langle r \rangle}{2} + \frac{c}{2\langle r \rangle} + \frac{d}{\langle r^2 \rangle}\right). \tag{37}$$

***Table 4.*** *$\langle r \rangle$ and $\langle v^2 \rangle$ of charmonium.*

| State | Parameter c | $\langle r \rangle(GeV^{-1})$ | $\langle v^2 \rangle$ |
|---|---|---|---|
| 1 S | 0.31384 | 1.92740 | 0.270037 |
| 1 P | 0.62740 | 2.61297 | 0.390517 |
| 1 D | 0.94121 | 3.16224 | 0.476976 |
| 2 S | 1.36707 | 2.76119 | 0.495803 |
| 2 P | 2.04345 | 3.32003 | 0.593240 |
| 3 S | 0.56582 | 3.10281 | 0.434766 |

**Table 5.** $\langle r \rangle$ and $\langle v^2 \rangle$ of bottomonium.

| State | Parameter c | $\langle r \rangle$(GeV$^{-1}$) | $\langle v^2 \rangle$ |
|---|---|---|---|
| 1 S | 0.51071 | 0.70781 | 0.109020 |
| 1 P | 1.02557 | 1.13682 | 0.153330 |
| 1 D | 1.53909 | 1.48158 | 0.190298 |
| 2 S | 1.19244 | 1.18887 | 0.167441 |
| 2 P | 1.82675 | 1.54697 | 0.209862 |
| 3 S | 0.95280 | 1.32665 | 0.146140 |

**Table 6.** $\langle r \rangle$ and $\langle v^2 \rangle$ of $b\bar{c}$ meson.

| State | Parameter c | $\langle r \rangle$(GeV$^{-1}$) | $\langle v^2 \rangle$ |
|---|---|---|---|
| 1 S | 1.35685 | 0.67088 | 0.469706 |
| 1 P | 2.71809 | 1.07872 | 0.603684 |
| 1 D | 3.15324 | 1.12857 | 0.658530 |
| 2 S | 4.82411 | 1.46948 | 0.763969 |
| 2 P | 2.53006 | 1.25885 | 0.532638 |

In Tables (4, 5, and 6). The quantities $\langle r \rangle$ and $\langle v^2 \rangle$ are computed for different states of charmonium meson, bottomonium meson and $b\bar{c}$-meson from Eqs. (36) and (37). The radius of $b\bar{b}$ is larger than the radius of $b\bar{c}$ and smaller than the radius of $c\bar{c}$. In addition, the values of radii of $b\bar{b}$, $b\bar{c}$ and $c\bar{c}$ are located in the interval 0.1 to 1 fm. This in agreement with Ref. [36].

## 5. Summary and Conclusion

In the present study, we employ the analytical exact iteration method (AEIM) for determining the energy eigenvalues and the wave functions of the multi-dimensional radial Schrödinger equation with the extended Cornell potential. The charmonium mass, bottomonium mass and $b\bar{c}$ meson mass are analytically obtained in the *N*-dimensional space and the special cases are obtained in comparison with other studies [2, 4, 27]. The effect of dimensional number is studied on the mass spectra of charmonium, bottomonium and $b\bar{c}$ meson in Tables (1, 2, and 3). Increasing dimensional number increases charmonium mass, bottomonium mass, and $b\bar{c}$ meson mass. The obtained results are in good agreement in comparison with Refs [2, 4, 27] and are in agreement with the experimental data. In addition, the mean radius $\langle r \rangle$ and the mean square velocity $\langle v^2 \rangle$ of the bound states of charmonium meson, bottomonium meson and $b\bar{c}$ meson are computed in Tables (4, 5, and 6). The radius of $b\bar{b}$ is larger than the radius of $b\bar{c}$ and smaller than the radius of $c\bar{c}$. This observation refers to one of the characteristics of quarkonium that the heavy quarkonium have smaller radii.

## References

[1] S. M. Kuchin and N. V. Maksimenko, Univ. J. Phys. Appl. 7, 295 (2013).

[2] R. Kumar and F. Chand, Common. Theor. Phys. 59, 528 (2013).

[3] T. Das, Electronic Journal of Theoretical Physics 13, No. 35, 207 (2016).

[4] A. Al-Jamel and H. Widyan, Appl. Phys. Rese. 4, 94 (2013).

[5] A. N. Ikot, O. A. Awoga, and A. D. Antia, Chin. Phys. 22, 020304 (2013).

[6] S. M. Al-Jaber, Int. J. Theor. Phys. 37, (1998).

[7] S. M. Al-Jaber, Int. J. Theor. Phys. 47, 1853 (2008).

[8] G. Chen, Z. Naturforsch. 59, 875 (2004).

[9] K. J. Oyewumi, F. O. Akinpelu and A. D. Agboola, Int. J. Theor. Phys. 47, 1039 (2008).

[10] S. M. Ikhdair and R. Sever, to appear in the Int. J. Mod. Phys. E (preprint quantph/0611065).

[11] D. Agboola, Phys. Scr. 80, 065304 (2009).

[12] H. Hassanabadi, Y. B. Hoda and LU. Liang-Liang, Chin. Phys. Lett. 29, 020303 (2012).

[13] S. Ikhdair, R. Sever, Journal of Molecular Structure: THEOCHEM 855, 13 (2008).

[14] H. Hassanabadi, S. Zarrinkamar and A. A. Rajabi, Commun. Theor. Phys. 55, 541 (2011).

[15] G. R. Khan, Eur. Phys. J. D 53, 123 (2009).

[16] H. Hassanabadi, B. H. Yazarloo, S. Zarrinkamar and M. Solaimani, Int. J. of Quantum Chemistry 112, 3706 (2012).

[17] G. R. Boroun and H. Abdolmalki, Phys. Scr. 80, 065003 (2009).

[18] M. Abu-shady, Journal of the Egyptian Mathematical Society accepted 22 June (2016).

[19] S. M. Ikhdair, M. Hamzavi, Physica B 407, 4797 (2012).

[20] E. Z. Liverts, E. G. Drukarev, R. Krivec, V. B. Mandelzweig, Few Body Syst. 44, 367 (2008).

[21] R. N. Choudhury, M. Mondal, Phys. Rep. A 52, 1850 (1995).

[22] R. De, R. Dutt, U. Sukhatme, J. Phys. A:Math.Gen. 25, L843 (1992).

[23] L. Gr. Ixaru, H. De. Meyer, G. V. Berghe, Phys. Rev. E 61, 3151 (2000).

[24] P. Sandin, M. Ögren, and M. Gulliksson, Phys. Rev. E 93, 033301 (2016)

[25] Z. Wang, Q. Chen, Comput. Phys. Commun. 179, 49 (2005).

[26] Chaudhuri R N and Mondal M Phys. Rev. A 52, 1850 (1995).

[27] R. Kumar, F. Chand, Phys. Scr. 85, 055008 (2012).

[28] H. Rahimov, H. Nikoofard, S. Zarrinkamar and H. Hassanabadi, Appl. Math. and Comp. 219, 4710 (2013).

[29] A. O. Barut, M. Berrondo, G. Garcia-Calderon, J. Math. Phys.

21, 1851(1980).

[30] S. özcelik, M. Simsek, Phys. Lett. A 152, 145 (1991).

[31] M. Abu-Shady, Inter. J. of Appl. Math. and Theor. Phys. 2, 16 (2016).

[32] A. Kumar Ray and P. C. Vinodkumar, Pramana J. Phys. 66, 958 (2006).

[33] E. J. Eichten and C. Quigg, Phys. Rev. D 49, 5845 (1994).

[34] D. Ebert, R. N. Faustov, and V. O. Galkin, Phys. Rev. D 67, 014027 (2003).

[35] N. V. Masksimenko and S. M. Kuchin, Russ. Phy. J. 54, 57 (2011).

[36] P. Gupta and I. Mehrotra, Journal of Modern Physics, 3, 1530 (2012).